\documentclass[amsmat,amssymb,amsfonts,aps,prb,twocolumn,showpacs]{revtex4}
\usepackage{graphicx}
\usepackage{dcolumn}
\usepackage{bm}

\usepackage{xcolor}

\usepackage{hyperref}
\usepackage{chemformula}
\usepackage{bbold}
\usepackage{amsmath}
\usepackage{siunitx}

\usepackage{bibunits}

\newcommand{\up}{\uparrow}
\newcommand{\dn}{\downarrow}

\begin{document}

\title{Topological magnons in CrI$_3$ monolayers: an itinerant fermion description}

\author{A. T. Costa $^{1}$,  
D. L. R. Santos$^{3}$,  
N. M. R. Peres$^{1,2}$, 
and J. Fern\'{a}ndez-Rossier$^{1,}$\footnote{On leave from Departamento de 
F\'{i}sica Aplicada, Universidad de Alicante, 03690 San Vicente del Raspeig, Spain.}} 
\affiliation{$^1$QuantaLab, International Iberian Nanotechnology Laboratory, 
4715-330 Braga, Portugal \\
$^2$Centro de F\'{i}sica das Universidades do Minho e Porto and Departamento de 
F\'{i}sica and QuantaLab, Universidade do Minho, Campus de Gualtar, 4710-057 Braga, Portugal\\
$^3$ Centro Federal de Educa\c c\~ao Tecnol\'ogica, Itagua\'i 23812-101, Rio de Janeiro, Brazil}

\date{\today}
\maketitle

\begin{bibunit}[apsrev4-1]

\textbf{
Magnons dominate the magnetic response of the recently discovered\cite{Huang2017} insulating 
ferromagnetic two dimensional crystals such as CrI$_3$. Because of the  
arrangement of the Cr spins in a honeycomb lattice, magnons in CrI$_3$ bear 
a strong resemblance with electronic quasiparticles in graphene\cite{Lado2017}. Neutron 
scattering experiments carried out in bulk CrI$_3$ show the existence of a 
gap at the Dirac points, that has been conjectured to have a topological 
nature\cite{Chen2018}. Here we propose a theory for magnons in ferromagnetic 
CrI$_3$ monolayers based on an itinerant fermion picture\cite{izuyama63}, with a Hamiltonian 
derived from first principles. We obtain the magnon dispersion for 2D CrI$_3$ 
with a gap at  the Dirac points with the same Berry curvature in both 
valleys. For CrI$_3$ ribbons, we find  chiral in-gap edge states. Analysis 
of the  magnon wave functions in momentum space further confirms their 
topological nature. Importantly, our approach does not require to define 
a spin Hamiltonian, and can be applied to both insulating and conducting 2D 
materials with any type of magnetic order.
}

Magnons are the Goldstone modes associated to the breaking of spin rotational 
symmetry. Therefore, they are the lowest energy excitations of magnetically 
ordered systems, and  their  contribution to thermodynamic properties, such 
as magnetization and specific heat, has been long acknowledged\cite{bloch1930,Herring51}. 
More recently, their  role in non-local spin current transport through magnetic 
insulators has been explored experimentally\cite{uchida10} and there are various 
proposals to use them for information processing in low dissipation spintronics\cite{chumak15}.  
In this context, the prediction of topological magnons with chiral edge modes\cite{shindou13,roldan2016,owerre2016} opens new horizons in 
the emerging field of topological magnonics\cite{wang18}. 
  
The recent discovery of  stand-alone 2D crystals with ferromagnetic order
down to the monolayer, such as CrI$_3$\cite{Huang2017}, CrGe$_2$Te$_6$
\cite{gong2017}, and others\cite{Gong19},  brings magnons to the center of the
stage, because of their even more prominent role determining the properties of
low dimensional magnets. Actually, an infinite number of magnons would be 
created  at any finite temperature in 2D magnets, unless magnetic
anisotropy or an applied magnetic field breaks spin rotational invariance and opens up a gap
at zero momentum\cite{mermin66,Lado2017}.
Unlike in 3D magnets, the thermodynamic properties of 2D magnets are
dramatically affected by the proliferation of magnons. This is the
ultimate reason of the very large dependence of the magnetization on the
magnetic field in materials with very small magnetic anisotropy, such as
CrGe$_2$Te$_6$\cite{gong2017}.

Magnons in CrI$_3$ attract strong
interest and are the subject of some controversy. Experimental probes include
inelastic electron tunneling\cite{Klein2018} and Raman spectroscopy\cite{jin18,cenker20} . In the case of bulk 
CrI$_3$, there are also 
ferromagnetic resonance \cite{lee20}  and inelastic neutron scattering experiments\cite{Chen2018}. Only the latter can provide access
to the full dispersion curves $E(k)=\hbar\omega(k)$.  There is a consensus that there are
two magnon branches, expected in a honeycomb lattice with two magnetic atoms
per unit cell. The lower branch has a finite minimum energy, $\Delta_\Gamma$, at the
zone center $\Gamma$. This energy represents the minimal energy cost to create a magnon
and plays thereby a crucial role. Different experiments provide radically
different values for $\Delta_\Gamma$, ranging from a fraction of meV to 9~meV\cite{jin18}.  This
quantity is related to the crystalline magnetic anisotropy energy that, according both  to 
Density Functional Theory (DFT) calculations\cite{Lado2017,torelli19} and multi-reference methods\cite{pizzochero19}, is in the range of 1~meV. 
\begin{figure*}
\center
\includegraphics[width=2\columnwidth]{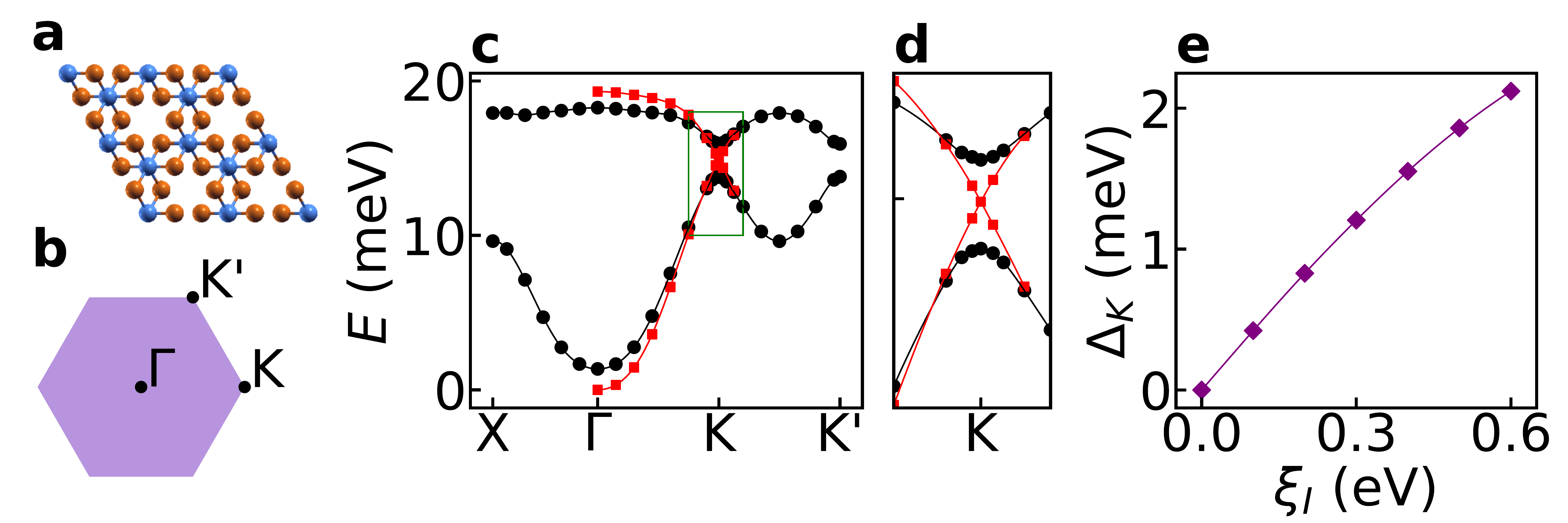}
\caption{{\bf Energy dispersion  magnons in CrI$_3$ monolayer.}   
\textbf{a}, Crystal structure of CrI$_3$  (top view, Cr atoms in blue, I atoms in orange).  
\textbf{b}, Brillouin zone, with high symmetry points.
\textbf{c}, Energy dispersion of magnons for 2D CrI$_3$ monolayer, 
obtained from the poles of the spin susceptibility tensor, computed 
for a ferromagnetic ground state (black circles). The data for the red squares were obtained from
a calculation where the SOC strength at the I atoms has been set to zero. 
({\bf d}) Zoom into the topological gap (the region marked by a green rectangle in panel \textbf{c}). 
In the absence of SOC the magnon  modes are degenerate at $K$, as evidenced by the red squares. 
({\bf e}) Size of the topological gap as a function of the SOC strength 
in the iodine atoms.}
\label{fig:dispersions_monolayer}
\end{figure*}

Inelastic Neutron scattering  also shows\cite{Chen2018} that, for bulk CrI$_3$,  the two
branches of the magnon dispersions are separated by a gap.  The minimum energy
splitting occurs at the $K$ and $K'$ points of the magnon Brillouin zone. As
in the case of other excitations in a honeycomb lattice with inversion
symmetry, such as electrons and phonons, one could expect a degeneracy of the
two branches at the Dirac cone, giving rise to Dirac magnons\cite{Pershoguba18}. 
Interestingly, second neighbour Dzyaloshinskii-Moriya (DM) interactions
are not forbbiden by symmetry in the CrI$_3$ honeycomb lattice, and are known to
open a topological gap\cite{owerre2016}, on account of mapping of ferromagneitc 
magnons with second neighbour DM in the honeycomb lattice into the 
Haldane Hamitonian \cite{Haldane}.

The description of magnons in magnetic 2D crystals has been exclusively
based in the definition of generalized Heisenberg spin Hamiltonians with various
anisotropy terms, such as single ion and XXZ exchange\cite{Lado2017}, Kitaev\cite{Kitaev},
DM\cite{owerre2016}. Once a given Hamiltonian is defined, the calculation of the spin
waves is relatively straightforward, using linear spin wave theory based on
Holstein-Primakoff representation of the spin operators\cite{holstein}. The
energy scales associated to these terms can be obtained both from fitting to
DFT calculations of magnetic configurations with various spin
arrangements\cite{Lado2017} as well as to some experiments\cite{Chen2018}.  
However, this method faces two severe limitations. First,
the symmetry and range of the interactions that have to be included in the spin Hamiltonian.
 are not clear a priori.
Second,  in order to determine $N$ energy constants,  $N+1$  DFT calculations forcing a ground state with a different 
magnetic arrangement are necessary and  the values so obtained can depend on the ansatz for the Hamiltonian.

\begin{figure*}
\center
\includegraphics[width=2\columnwidth]{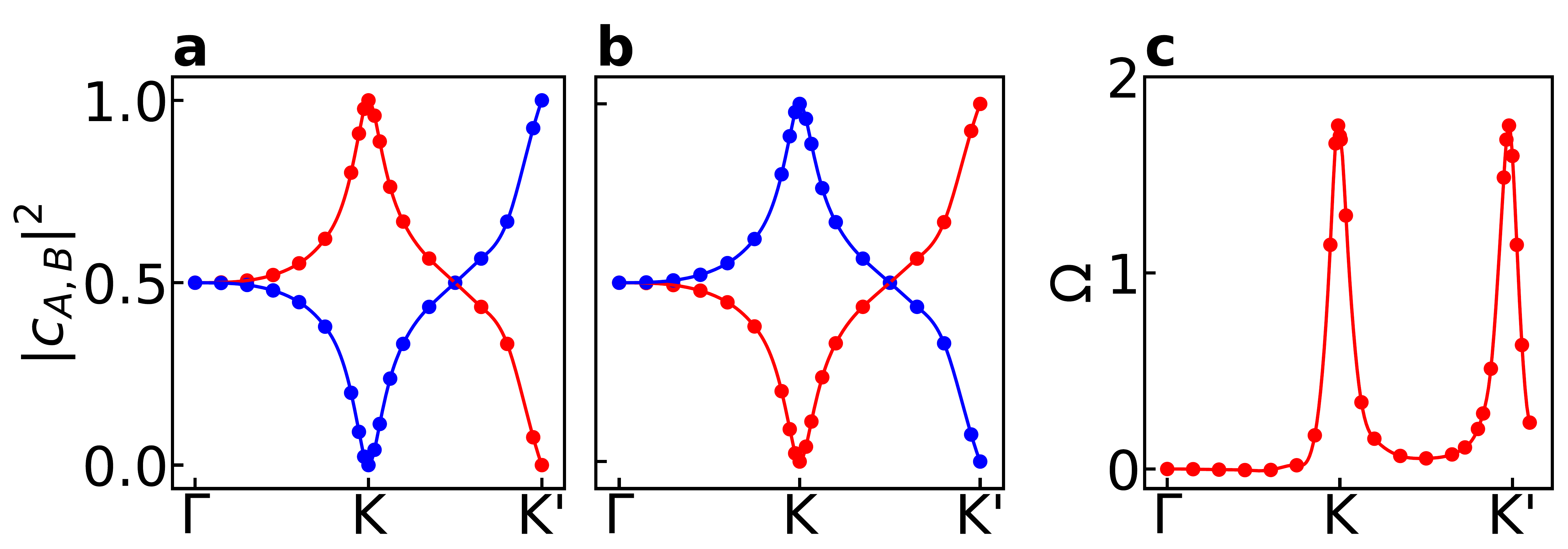}
\caption{\textbf{Analysis of the  magnon wave function coefficients  
(eq. \ref{magnoncoeff}) for a CrI$_3$ monolayer}.  
Coefficients $c_A$ and $c_B$ for lower (\textbf{a}) and higher (\textbf{b}) 
energy branch along the $\Gamma,K,K'$ line in the Brillouin zone.  
It is apparent that at the Dirac points $K,K'$, the spinor is 
sublattice polarized: the sign of the polarization changes as we 
change either the branch or the mode, following a braiding pattern, 
exactly like in the Haldane model.
\textbf{c}, Berry curvature, for both magnon branches, along the $\Gamma,K,K'$ 
line in the Brillouin zone. For a given branch, the Berry curvature has 
the same sign in both valleys, that give the dominant contribution.
The sign of the Berry curvature is opposite for both branches. Thus, 
the integrated Berry curvature is clearly finite, with opposite signs 
for the 2 branches.}
\label{fig:psibulk}
\end{figure*}

Here we circumvent this methodological bottleneck and describe magnons directly from 
an itinerant fermion model derived from first principles calculations. Our approach, that has been extensively used to describe magnons in itinerant magnets\cite{antc:2006:CoCuinterm,antc:2010:SOCMethod}, is carried out in five steps. 
First, we compute the electronic structure of the material using DFT 
in the generalized gradient approximation (GGA), without taking either 
spin polarization or spin orbit coupling (SOC) into account. Second, we derive a 
tight-binding model with $s$,$p$, $d$ shells in Cr and $s$ and $p$ shells in Iodine. 
The electronic
bands obtained from this Hamiltonian are identical to those calculated from  DFT  (see methods for details). In the third step we include both SOC in Cr and I  as well as 
 on-site intra-atomic  Coulomb repulsion in the Cr $d$. The resulting model is solved 
 in a self-consistent mean field approximation\cite{antc:2010:SOCMethod}.  The strength of the Coulomb repulsion is chosen to reproduce the DFT 
magnetic moment per Cr atom, and the strength of the spin-orbit coupling is taken 
from the literature\cite{handbookphotochem}.
 In the  fourth step we compute the generalized 
spin susceptibility tensor $\pmb{\chi}(\vec{q},\omega)$ in the random phase 
approximation (RPA). In the final step we find the poles of the spin suceptibility tensor 
in the $(\omega, \vec{q})$ space, that define the dispersion relation 
$E_n(\vec{q})=\hbar\omega_{n}(\vec{q})$ of the magnon modes, where $n$ labels the different modes.

\begin {figure*}[t]
\center
\includegraphics[width=2\columnwidth]{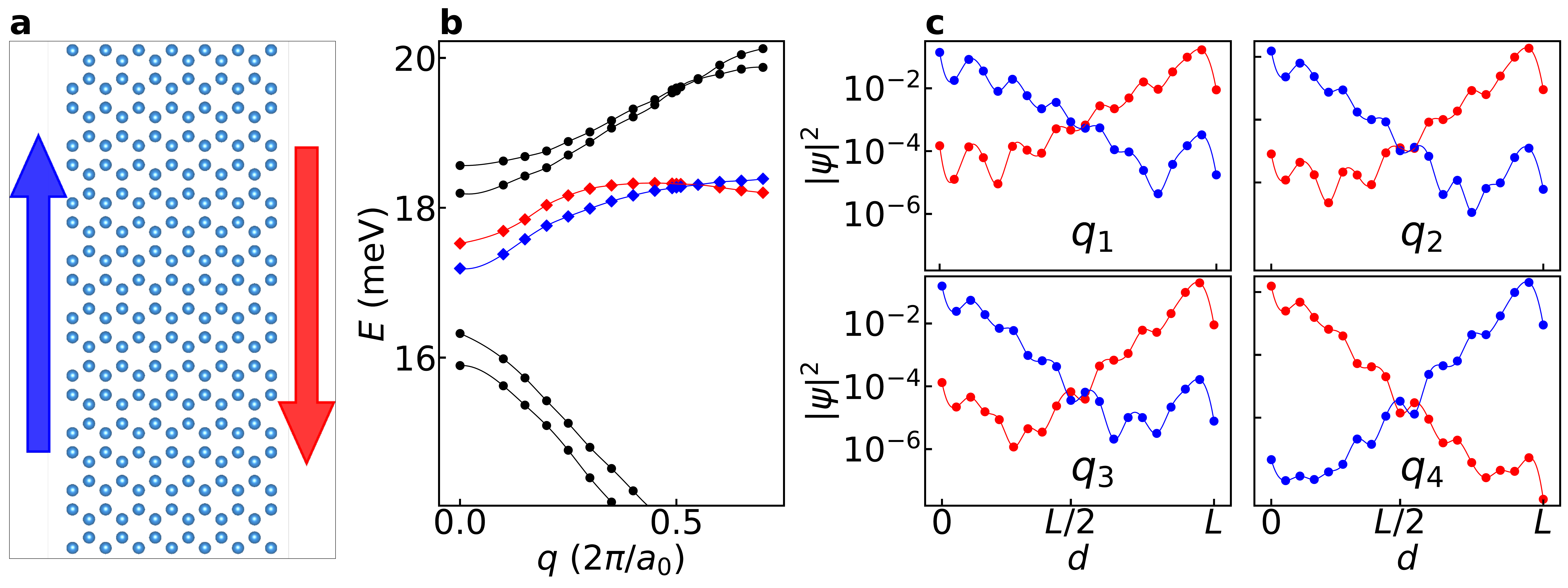}
\caption{\textbf{Magnons in CrI$_3$ nanoribbon.} \textbf{a}, sketch of the Cr sites of the ribbon unit cell. The iodine sites are not shown, for clarity. \textbf{b}, Dispersion of the ribbon magnons, zoomed at the energy of the Dirac gap.  The data highlighted in red and blue belong to the edge modes. \textbf{c}, Probability density for the edge modes as a function of the distance $d$ from the left edge of the ribbon, for various wave vectors: $q_1=0.40$,
$q_2=0.50$, $q_3=0.55$, $q_4=0.60$, in units of $2\pi/a_0$. The colors match those for the corresponding branches in \textbf{b}.
}
\label{fig:ribbon}
\end{figure*}

 The   2D CrI$_3$  magnon dispersion along the high symmetry directions 
 of the  Brillouin zone (BZ) are shown in figure~\ref{fig:dispersions_monolayer}, calculated both with and without spin orbit coupling, $\xi_I$.
 As expected for a unit cell with two magnetic atoms, we find two branches of magnons.  
 At the $\Gamma$ point, spin orbit coupling opens up a gap $\Delta_{\Gamma}$, 
 as expected\cite{Lado2017}. At the $K$ and $K'$ points, the two magnon branches 
 form Dirac cones when $\xi_I=0$,  but a gap $\Delta_{K,K'}$ opens up, 
 whose magnitude is an increasing function of the iodine spin orbit coupling. 
 
In order to assess  the topological nature of the gap at $K$ and $K'$ points, we first examine the wave functions for the two modes along the $\Gamma-K-K'$ line. 
The magnon wave functions can be written as linear combinations of spin flips
across the Cr honeycomb lattice, with weights $c_A$ and $c_B$ on the $A$ and $B$ 
triangular sublattices:
\begin{equation}
|\Psi_n(\vec{q})\rangle= \sum_{\vec{R}}
\left[ c_A(n,\vec{q}) |A,\vec{R}\rangle +c_B(n,\vec{q}) |B,\vec{R}\rangle\right]e^{i\vec{q}\cdot\vec{R}}
\label{magnoncoeff}
\end{equation}
where $n$ labels the branch. 
A distinctive feature of  topological quasiparticles in the honeycomb lattice\cite{Haldane,KaneMele}
is the {\em braiding} in momentum space of the sublattice components. 
In panels a and b of figure~\ref{fig:psibulk} we plot the coefficients $c_A(n,\vec{q})$ and $c_B(n,\vec{q})$, obtained from our itinerant fermion model, as $\vec{q}$ traces the high symmetry directions of the magnon Brillouin zone. As $\vec{q}$ goes from $K$ to $K'$,  for a given $n$, $(c_A,c_B)$ behaves as a spinor that goes 
from the north to the south pole, with the reverse behaviour for the other branch, exactly as in the Haldane model.  
  We have verified that this pattern is reversed if the off-plane magnetization changes sign. 

Topological magnons have a finite Berry curvature that leads to non-zero Chern
number when integrated over the entire BZ\cite{shindou13,roldan2016,owerre2016}. 
In figure~\ref{fig:psibulk}c we show the Berry curvature along the high symmetry line $\Gamma$-$K$-$K'$
in the Brillouin zone (see supplementary material).  The  Berry curvature of a given mode peaks at the $K$ and $K'$ valleys, with the same sign. Therefore, we expect a non-zero Chern number and hence the existence of in-gap chiral edge modes. \footnote{Because of the very large computational cost, we have not 
tried a complete integration of the BZ.}

Our calculations (Fig. 1e) give strong evidence that  the topological gap is driven by the spin orbit coupling of iodine. Thus, the finite Berry curvature has to be produce by inter-atomic exchange mediated by the ligand. A very likely candidate is second neighbour DM interactions, that are known to result in topological magnons in honeycomb ferromagnets\cite{owerre2016}.
 
We now address the case of magnons in a CrI$_3$ ribbon, using the itinerant fermion description, in order to look for
topological edge states.  
 We consider a ribbon where the edge Cr atoms form an armchair pattern, to avoid non-topological modes that arise at  zigzag edges. 
 The unit cell used in the calculations has 40 Cr atoms, wide enough to prevent cross-talk between edges. Therefore, for a given value of the longitudinal wave vector $q$, there are 40 magnon modes.
In order to avoid an extremely heavy calculation, we use the bulk
fermionic  tight-binding parameters for the ribbon, neglecting thereby changes in the electronic structure that may arise at the edges.  As a result, the obtained  value of $\Delta_{\Gamma}$ for the ribbon is $\sim$~1 meV higher.

A zoom of the resulting energy dispersion, around the Dirac energy, is shown in figure~\ref{fig:ribbon}b. 
The red and blue diamonds indicate modes that are exponentially localized at either edge of
the ribbon, as shown in figure~\ref{fig:ribbon}c. 
 Our results strongly indicate  the existence of localized  modes at   the CrI$_3$ edges, across the entire one dimensional BZ.  Around the Dirac point these edge modes  are chiral, and their energy is inside the gap. Away from the Dirac point their 
 dispersions are not linear due to the presence of long range exchange.

The topological nature of  CrI$_3$ magnons entails two consequences. First, as a result of the finite Berry curvature,  magnons contribute to the  thermal Hall conductivity at zero magnetic field\cite{Katsura10}. Second, a specific consquence of the quantized Chern number is the existence of edge modes.  Our calculations show they 
have narrow spectral features. Therefore, their existence could be confirmed by  inelastic electron tunnel spectroscopy carried out with a scanning probe \cite{spinelli14} to determine the local density of states of spin excitations with atomic resolution.  
      
Our method to obtain the magnons  directly from a microscopic
electronic Hamiltonian derived from ab initio calculations is 
widely applicable to 2D materials and their heterostructures. 
The method can also be used to obtain spin excitations 
from non-collinear and non-coplanar ground states and to  examine 
the stability of competing  states, which can  prove extremely useful in unveiling
the nature of the magnetic ground state of Kitaev  materials such as $\alpha$-RuCl$_3$.
   
\putbib[biblio]
\end{bibunit}

\acknowledgments{
We  acknowledge useful discussions with M. Costa, R. B. Muniz, A. Molina-S\'anchez and D. Soriano.
 N. M. R. P.  acknowledges support from the European Commission through the project "Graphene- Driven Revolutions in ICT and Beyond" (Ref. No. 785219), and the Portuguese Foundation for Science and Technology (FCT) in the framework of the Strategic Financing UID/FIS/04650/2013,   COMPETE2020, PORTUGAL2020, FEDER and the Portuguese Foundation for Science and Technology (FCT) through projects PTDC/FIS-NAN/3668/2013 and POCI-01-0145-FEDER-028114.
J. F.-R. acknowledges financial support 
from FCT  UTAPEXPL/NTec/0046/2017 project, as well as Generalitat
Valenciana funding Prometeo2017/139 and MINECO Spain (Grant No. MAT2016-78625-C2).
D.L.R.S. thankfully acknowledges the use of HPC resources provided by the National Laboratory for Scientific Computing (LNCC/MCTI, Brazil). A.T.C. thankfully acknowledges the use of computer resources at MareNostrum and the technical support provided by Barcelona Supercomputing Center (RES-FI-2019-2-0034).}

\appendix

\clearpage

\begin{bibunit}[apsrev4-1]

\noindent
\textbf{\large Methods}

\noindent
\textbf{Density Functional Theory calculations}.
The DFT calculation 
has been performed with the Quantum Espresso package\cite{Giannozzi2009,Giannozzi2017}. 
We employed the PBE functional\cite{PhysRevLett.77.3865} and the ionic potentials were 
described through the use of projected augmented wave (PAW)
pseudopotentials\cite{PhysRevB.59.1758}. The energy cutoff 
for plane waves was set to 80~Ry. We used a $25\times25\times1$ Monkhorst-Pack 
reciprocal space mesh\cite{PhysRevB.13.5188}. 

\bigskip 

\noindent
\textbf{Fermionic Hamiltonian}
We  now describe steps 2 and 3 of the method outlined in the main text.
The electronic states of the CrI$_3$ monolayer are described by a model Hamiltonian
\begin{equation}
H = H_0 + H_I + H_\mathrm{SOC}
\end{equation}
The first term, describing the tight-binding Hamiltonian for $s,p,d$ orbitals in Cr and $s,p$ orbitals in I
is given by
\begin{equation}
H_0 = \sum_{ll'}\sum_{\mu\mu'}\sum_\sigma 
T^{\mu\mu'}_{ll'}a^\dagger_{l\mu\sigma}a_{l'\mu'\sigma},
\end{equation}
The hopping matrix  $T^{\mu\mu'}_{ll'}$ is extracted by the pseudo atomic orbital projection 
method\cite{PhysRevB.88.165127,PhysRevX.5.011006,PhysRevB.93.035104,PhysRevB.93.125137,BUONGIORNONARDELLI2018462}. The method consists in projecting the 
Hilbert space spanned by the plane waves onto a compact subspace composed of 
the pseudo atomic orbitals (PAO). These PAO functions are naturally built into 
the pseudo potential used in the DFT calculation. The bands obtained from this tight-binding model, are identical with those obtained from the spin un-polarized DFT calculation  (see 
suppl. mat. figure S1).

In the third step of the method, we add  both a screened Coulomb repulsion term,
\begin{equation}
H_I = \sum_{l}\sum_{\mu\mu'\nu\nu'}\sum_{\sigma\sigma'}
I^{\mu\mu'\nu\nu'}a^\dagger_{l\mu\sigma}a^\dagger_{l\mu'\sigma'}a_{l\nu'\sigma'}a_{l\nu\sigma},
\end{equation}
and a local spin-orbit coupling (SOC) term,
\begin{equation}
H_\mathrm{SOC} = \sum_{l}\sum_{\mu\mu'}\sum_{\sigma\sigma'}
\xi_l\langle R_l\mu\sigma|\vec{L}\cdot\vec{s}|R_l\mu'\sigma'\rangle
a^\dagger_{l\mu\sigma}a_{l\mu'\sigma'}.
\end{equation}

The screened Coulomb repulsion matrix elements $I^{\mu\mu'\nu\nu'}(R_l)$ are 
approximated by a single parameter form, which is qualitatively equivalent to taking 
a spherically symmetric average of the interaction potential~\cite{LowdeWindsor},
\begin{equation}
I^{\mu\mu'\nu\nu'}(R_l)=I(R_l)\delta_{\mu\nu'}\delta_{\mu'\nu}.
\end{equation}
We further assume the repulsion between electrons in $s$ and $p$ orbitals
is negligible. Thus, only electrons occupying $d$ orbitals at Cr atoms suffer
electron-electron repulsion.

The spin-polarized ground-state of the system is obtained within a
self-consistent mean-field approximation, in which all three components
of the magnetization of each Cr atom within the unit cell are treated as 
independent variables.

\bigskip

\noindent
\textbf{Fermionic Spin susceptibility in the RPA approximation.}
The magnon energies are associated with the poles of the frequency-dependent
transverse spin susceptibility,
\begin{equation}
\chi^{+-}(R_l,R_{l'},\omega) \equiv \int_{-\infty}^\infty dt e^{i\omega t} \chi^{+-}(R_l,R_{l'},t),
\end{equation}
where
\begin{equation}
\chi^{+-}(R_l,R_{l'},t) \equiv 
-i\theta(t-t')\left\langle\left[ S^+_l(t),S^-_{l'}(0)\right]\right\rangle ,
\label{magnonpropagator}
\end{equation}
and 
\begin{equation}
 S^+_l\equiv \sum_\mu a^\dagger_{l\mu\up}a_{l\mu\dn}.
\end{equation}
$a^\dagger_{l\mu\sigma}$ is the cration operator for an 
atomic-like orbital $\mu$ at site $R_l$ with spin $\sigma=\up,\dn$.
The angular brackets $\langle\cdots\rangle$ represent a thermal average
over the grand-canonical ensemble. The double time Green function $\chi^{+-}(R_l,R_{l'},t)$  
defined in equation~\ref{magnonpropagator} can be interpreted as the propagator for 
localized spin excitations created by the operator $S^-_{l'}$. In a system with
translation invariance, its reciprocal space counterpart can be readily interpreted as
the propagator for magnons with well-defined wave vector.

The transverse spin susceptibility is calculated within a time-dependent
mean-field approximation, which is equivalent to summing up all ladder 
diagrams in the perturbative series for $\chi^{+-}$. These are the same
Feynman diagrams that enter into time-dependent density functional theory.
In the presence of SOC, however, the transverse susceptibility becomes coupled
to other three susceptibilities, which are related to longitudinal 
fluctuations of the spin density and fluctuations of the charge density.
Thus, it becomes necessary to solve simultaneously the equations of motion 
for the four susceptibilities\cite{antc:2010:SOCMethod}. 

\bigskip

\noindent
\textbf{Berry curvature calculation.}
The Berry phase associated to a closed contour $C$
 in the momentum space $\vec{k}$ is given by\cite{Berry1984}:
\begin{equation}
\gamma_n= i \oint_C \vec{{\cal A}_n}(\vec{k}) \cdot d\vec{k} =
-{\rm Im}\int_S \vec{{\cal B}}(\vec{k}) dS
\end{equation} 
where
$
\vec{ {\cal A}_n}(\vec{k})\equiv  \langle \Psi_n(\vec{k})|\nabla_{\vec{k}}\Psi_n(\vec{k})\rangle
$
 is the Berry connection and
$\vec{{\cal B}_n}(\vec{k})= \vec{\nabla}\times {\cal A}(\vec{k})$
 is the Berry curvature.  
 
An efficient way to compute the Berry curvature at a given point $\vec{k}_0$ is to compute the Berry 
phase in a infinitesimal loop in the plane $(k_x,k_y)$\cite{asboth2016}. We  parametrize the line integral with the variable 
 $\theta$, 
\begin{equation}
\gamma_n=i\oint \langle \Psi_n(\theta)|\frac{\partial\Psi_n(\theta)}{\partial \theta}\rangle d\theta .
\end{equation}
Now we note that the argument of the integral has to be purely imaginary, since  $\frac{ \partial  \langle \Psi_n(\theta)|\Psi_n(\theta)\rangle }{\partial\theta}=0$.
We thus have:
\begin{equation}
\gamma_n=-{\rm Im}\oint \langle \Psi_n(\theta)|\frac{\partial\Psi_n(\theta)}{\partial \theta}\rangle d\theta .
\end{equation}
We discretize the integral and the derivative:
\begin{equation}
\gamma_n\simeq  -{\rm Im} \sum_{j=1,N}  \langle \Psi_n(\theta)
|\frac{\Psi_n(\theta_j+\Delta \theta)\rangle-\Psi_n(\theta_j)\rangle}{\Delta \theta}\rangle \Delta \theta
\end{equation}
We expand this expression:
\begin{equation}
\gamma_n\simeq  -{\rm Im} \sum_{j=1,N} \left( \langle \Psi_n(\theta)
|\Psi_n(\theta_j+\Delta \theta)\rangle-1 \right) .
\end{equation}
Now we use the fact that the overlap is close to 1 so that $\epsilon= \left( \langle \Psi_n(\theta)
|\Psi_n(\theta_j+\Delta \theta)\rangle-1 \right)$ is a small number. We use the expression $\log (1+\epsilon)\simeq \epsilon$  and write:
\begin{equation}
\gamma_n\simeq  i \sum_{j=1,N} \log \left( \langle \Psi_n(\theta)
|\Psi_n(\theta_j+\Delta \theta)\rangle \right) .
\end{equation}
Now we use  $\sum_i \log f_j = \log (\prod_j f_j)$ to write:
\begin{equation}
\gamma_n\simeq  -{\rm Im}  \log \prod_j \left( \langle \Psi_n(\theta)
|\Psi_n(\theta_j+\Delta \theta)\rangle \right) .
\end{equation}
This expression is convenient for numerical evaluation, because random phases are eliminated, as all states appear twice as conjugated pairs.
 Therefore, random phases that inevitably occur in the numerical diagonalizations are cancelled. 

We now consider an infinitesimal loop of area $\frac{1}{2}(\Delta k)^2$ formed by 3 points, $\vec{k}_0$, $\vec{k}_1=\vec{k}_0+(\Delta k,0)$, $\vec{k}_2=\vec{k}_0+(0,\Delta k)$.   
We now introduce the notation for the overlap  
\begin{equation}
O_{i,j}\equiv \langle \Psi_n(\vec{k}_i) |\Psi_n(\vec{k}_j)\rangle
\end{equation}
to write the Berry phase in the loop as
\begin{equation}
\gamma(\vec{k}_0) \simeq  -{\rm Im} \log \left(O_{0,1}O_{1,2} O_{2,0} 
\right)= {\cal B}(\vec{k}_0) \frac{1}{2}(\Delta k)^2 .
\end{equation}
Thus, the Berry curvature is obtained as:
\begin{equation}
{\cal B}(\vec{k}_0)=-\frac{2}{(\Delta k)^2}
{\rm Im} \log \left(O_{0,1}O_{1,2} O_{2,0} 
\right) .
\end{equation}

\putbib[biblio]
\end{bibunit}

\end{document}


\title{Supplementary Material for\\ 
Topological magnons in CrI$_3$ monolayers: an itinerant fermion description}

\author{A. T. Costa $^{1}$,  
D. L. R. Santos$^{3}$,  
N. M. R. Peres$^{1,2}$, 
and J. Fern\'{a}ndez-Rossier$^{1,}$\footnote{On leave from Departamento de F\'{i}sica Aplicada, Universidad de Alicante, 03690 San Vicente del Raspeig, Spain.}} 
\affiliation{$^1$QuantaLab, International Iberian Nanotechnology Laboratory, 4715-330 Braga, Portugal \\
$^2$Centro de F\'{i}sica das Universidades do Minho e Porto and Departamento de F\'{i}sica and QuantaLab, Universidade do Minho, Campus de Gualtar, 4710-057 Braga, Portugal\\
$^3$ Centro Federal de Educação Tecnológica, Itagua\'i 23812-101, Rio de Janeiro, Brazil}

\date{\today}

\begin{abstract} 
\end{abstract}

\maketitle

\section{DFT and LCAO bands}
\begin{figure}
\center
\includegraphics[width=\columnwidth]{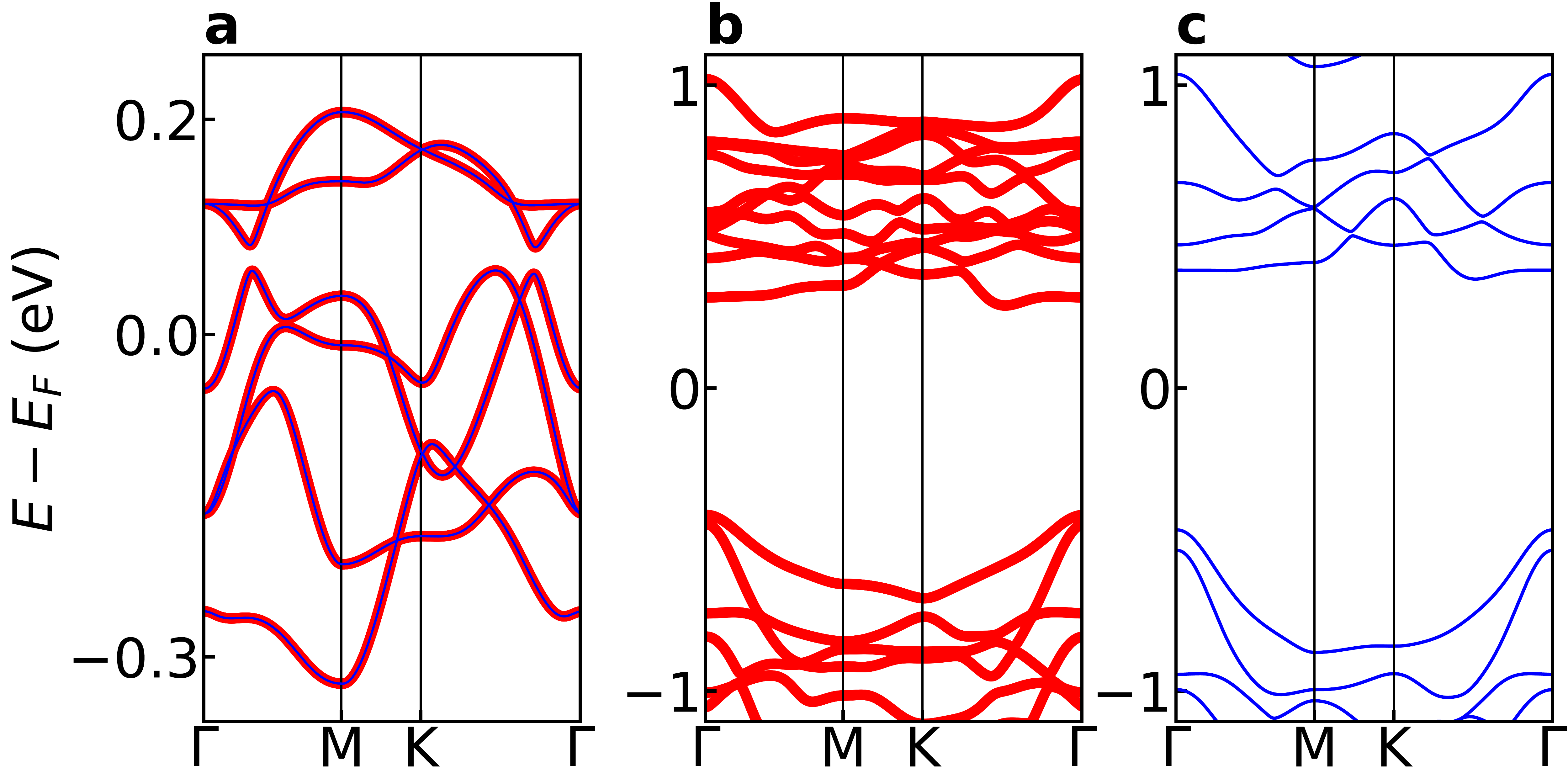}
\caption{{\bf Electronic band structure of a CrI$_3$ monolayer. }   
({\bf a}) The results of a DFT calculation without spin polarization
or SOC (solid blue lines) superimposed
to the results of a LCAO calculation with a hopping matrix derived from
the same DFT calculation. Spin-orbit coupling and spin polarization have 
been turned off for both calculations. $E_F$ stands for Fermi energy.
({\bf b}) LCAO bands after self-consistent mean-field calculation
including intra-atomic Coulomb repulsion and spin-orbit coupling.
({\bf c}) DFT bands with spin polarization and spin-orbit coupling.}
\label{fig:bands}
\end{figure}

Here we present the band structure of a CrI$_3$ monolayer as obtained from the same ab 
initio calculation from which we extracted the hopping matrix used our the susceptibility 
calculations. In the original DFT calculation (described in detail in the ``Methods'' section 
of the manuscript) spin polarization is suppressed and spin-orbit coupling is turned off.
The resulting band structure is shown in figure~\ref{fig:bands}a, together with the bands
obtained from the corresponding LCAO hamiltonian. We also show the LCAO band structure after
the inclusion of Coulomb repulsion (leading to spin polarization) and spin-orbit coupling 
(figure~\ref{fig:bands}b). For comparison, we show in panel c of figure~\ref{fig:bands}
the bands obtained from a DFT calculation with SOC and spin polarization.

\section{Magnon normal modes}
In a system lacking periodicity, or with more than one magnetic atom per unit cell,
the frequency- (and eventually wave vector-) dependent transverse spin susceptibility 
can be written as a matrix in atomic site indices, $\chi^{\perp}_{ll'}(\omega)$. There are at least
two useful interpretations for this matrix. One originates from its role as a response function
in the linear regime, the other is related to its formal similarity to the single-particle
Green function of many-body theory.

\subsection{ $\chi^{\perp}_{ll'}(\omega)$ as a linear response function}
When interpreted as a linear response function the transverse spin
susceptibility yields the change in the transverse component of the spin
moment $\delta S^+_l$ at site $l$ due to a transverse, circularly polarized external 
field $b_{l'}$ of frequency $\omega$ acting on site $l'$,
\begin{equation}
\delta S^+_l = \sum_{l'}\chi^{\perp}_{ll'}b_{l'}.
\end{equation}
We assume the system has $N$ magnon normal modes, where $N$ equals the number
of non-equivalent magnetic atoms in the system. Each mode $(m)$ is characterized 
by complex amplitudes $\xi^{(m)}_l$ at the magnetic site $l$. A general motion of the 
transverse components of the spin can be written as a linear combination of the normal
modes,
\begin{equation}
\delta S^+_l = \sum_m \psi_m\xi^{(m)}_l .
\end{equation}
Now consider an external field whose frequency and complex amplitudes match exactly those
of a normal mode,
\begin{equation}
b_l = b_0\xi^{(m)}_l .
\end{equation}
In this case, the corresponding change in the transverse spin moment $\delta S^+_l$
induced by the field should be proportional to the the same normal mode,
\begin{equation}
\delta S^+_l = \sum_{l'}\chi^{\perp}_{ll'}(\omega)b_{l'} = s_0\xi^{(m)}_l .
\end{equation}
Thus,
\begin{equation}
\sum_{l'}\chi^{\perp}_{ll'}(\omega)\xi^{(m)}_l  = \frac{s_0}{b_0}\xi^{(m)}_l .
\label{eq:eigen}
\end{equation}
This shows that the normal modes are the eigenvectors of the susceptibility matrix.
In principle, this procedure yields ``normal modes'' for any arbitrary frequency
of the external field. However, the ``true'' normal modes are the ones for which
the system responds resonantly. Thus, we can look at the imaginary part of the 
eigenvalues of $\chi^{\perp}_{ll'}(\omega)$ as a function of frequency and associate 
their peaks with the frequencies of the normal modes. 

\subsection{ $\chi^{\perp}_{ll'}(\omega)$ as the magnon singe-particle Green function}

In order to arrive at this interpretation we can make an analogy with the spin wave theory obtained
from the linearized Holstein-Primakoff transformation~\cite{PhysRev.58.1098}. There, after 
linearization, the bosonic operator that represents a spin excitation localized at atomic site $l$
is $b^\dagger\equiv S_l^-$. Thus, if we write the definition of the transverse susceptibility replacing $S_l^+$by $b_l$ and
$S_l^-$ by $b^\dagger$, we arrive at a form that is completely analogous to
that of the single particle Green function of many-body theory,
\begin{equation}
\chi^{\perp}_{ll'}(\omega) \equiv -i\theta(t)\left\langle\left[ b_l(t),b_{l'}^\dagger(0)\right]\right\rangle .
\end{equation}
Here, $\langle\cdots\rangle$ is a thermal average (or a ground state average at $T=0$), $\theta(t)$ is the Heaviside
unit step function. As in the linearized HP transformation, the magnons of our RPA theory are
independent particles, described by an effective hamiltonian $H$ composed only of one-body terms. In that
case, it is straightforward to show that the Fourier transform of the single-particle Green functions $\chi_{ll'}$ are the matrix elements of a matrix $\pmb{\chi}$ related to the hamiltonian matrix  by
\begin{equation}
    \pmb{\chi}(E) = (E-\mathbf{H})^{-1}.
\end{equation}
Thus, the magnon normal modes of the system are the eigenvectors of the susceptibility matrix $\pmb{\chi}(E^*)$
where $E^*$ are the magnon energies, associated with the peaks of the imaginary part of the eigenvalues of $\pmb{\chi}$.

\bibliographystyle{apsrev4-1}
\bibliography{biblio}{}